# Power-Efficient Ultra-Wideband Waveform Design Considering Radio Channel Effects

Yipeng Liu[*], Qun Wan[*], Xiaoli Chu[†]

[*] Department of Electronic Engineering, University of Electronic Science and Technology of China (UESTC), Chengdu, 611731, China

[†] Division of Engineering, King's College London, London WC2R 2LS, UK

liuyipeng@uestc.edu.cn; wanqun@uestc.edu.cn; xiaoli.chu@kcl.ac.uk

*Abstract* —This paper presents a power-efficient mask-constrained ultra-wideband (UWB) waveform design with radio channel effects taken into consideration. Based on a finite impulse response (FIR) filter, we develop a convex optimization model with respect to the autocorrelation of the filter coefficients to optimize the transmitted signal power spectrum, subject to a regulatory emission mask. To improve power efficiency, effects of transmitter radio frequency (RF) components are included in the optimization of the transmitter-output waveform, and radio propagation effects are considered for optimizing at the receiver. Optimum coefficients of the FIR filter are obtained through spectral factorization of their autocorrelations. Simulation results show that the proposed method is able to maximize the transmitted UWB signal power under mask constraints set by regulatory authorities, while mitigating the power loss caused by channel attenuations.

*Index Terms* —power efficiency, ultra-wideband, green communication, radio propagation effect, waveform design.

I. INTRODUCTION

Power-efficient transmissions play a key role in the emerging green communications [1], which is becoming increasingly important in meeting demands for environmental sustainability. Ultra-wideband (UWB) has a number of unique merits [2], making it a promising technology for high-data-rate short-range wireless communications. The low-power merit of UWB also makes it an attractive candidate for realizing green communications.

As UWB signals occupy enormous bandwidths, the UWB transmission power has to be confined to a sufficiently low level [3] to avoid interference with other wireless systems operating on the same spectrum. Therefore, the UWB emission mask that is imposed by regulatory authorities must be taken into account for achieving power-efficient UWB communications. At the same time, UWB signals may suffer distortions caused by radio frequency (RF) components of the transmitter and the frequency selective UWB propagation channel [4]. Waveform distortions and radio propagation losses are two of the major factors that deteriorate UWB communications performance. In this work, we use the US FCC UWB spectrum mask [3] as a reference mask, but the proposed algorithm can be generalized to other regulatory masks for UWB communications.

Existing designs of UWB waveforms do not fit the emission mask efficiently. By sampling the given spectrum mask, an UWB waveform design algorithm based on eigenvalue decomposition was proposed in [5]. In [6], a Parks-McCllan (PM) algorithm was used to get an approximation of the spectrum mask in a minimax sense. However, these two algorithms do not directly optimize the spectral utilization. In [7], based on a finite impulse response (FIR) filter, a convex optimization model with respect to the autocorrelation of the filter's coefficients was formed to get the optimal mask utility ratio which was defined as the ratio of the power of the synthesized UWB waveform to the total power permissible under the spectrum mask. and the optimal filter coefficients were obtained through spectral factorization of the autocorrelation.

On the other hand, most existing UWB waveform design algorithms [5]-[7] did not incorporate the effects of transmitter RF components, such as amplifiers, samplers and antennas, and other radio propagation effects, but these effects may significantly distort the designed waveform, leading to a transmitted power spectrum that actually does not match the given spectrum mask. A subsequent shift of the transmitted power spectrum back under the spectrum mask may result in a considerable loss in power efficiency. In order to improve power efficiency, based on the method in [7], this paper proposed a UWB waveform considering effects of transmitter RF components so that the transmitted waveform have the optimal mask utility ratio, and radio propagation effects are take into account too.

In the rest of this paper, Section II introduces the UWB signal model with the FIR filter for waveform shaping. The power-efficient UWB waveform design is proposed in Section III. Simulation results are presented in Section IV to evaluate the performance of the proposed algorithm. Finally conclusions are given in Section V.

## II. Signal Model

In pulse-position and/or pulse-amplitude modulated UWB impulse radio systems, when elements of the time hopping sequences are independent and identically distributed integer-valued random variables and when the polarity randomization is applied, the power spectral density (PSD) of the UWB signal is given by $\Phi(f) = \alpha |P(f)|^2$ [8], where $\alpha$ is a constant and $P(f)$ is the frequency response of the UWB waveform $p(t)$. The UWB waveform design is equivalent to the design of the waveform function $p(t)$.

We adopt an FIR filter to generate the basic waveform $p(t)$, which can be formulated as

$$p(t) = \sum_{k=0}^{L-1} g_k q(t - kT_0), \tag{1}$$

where $g_k$ are real-valued filter tap coefficients, $L$ is the total number of taps, $q(t)$ is an elementary pulse with a duration of $T_q$, $T_0$ is the sampling interval, and the pulse duration $T_p$ of $p(t)$ is given by $T_p = (L-1)T_0 + T_q$. The clock rate of the transmitter is $F_0 = 1/T_0$.

III. WAVEFORM DESIGN

*A. Maximization of Transmitted Waveform Power*

At the transmitter, the UWB waveform $p(t)$ is synthesized as in (1). Its corresponding power spectrum is given by

$$S_p(f) = S_g(f) S_q(f), \tag{2}$$

where $S_g(f)$ and $S_q(f)$ are the power spectra of $(g_k)$ and $q(t)$, respectively.

We define the Fourier transformation vector as

$$\mathbf{v}(f, L) = \left[1, e^{j2\pi f T_0}, e^{j2\pi f 2T_0}, \cdots, e^{j2\pi f (L-1)T_0}\right]^T, \tag{3}$$

and define the filter coefficients vector as

$$\mathbf{g} = \begin{bmatrix} g_0 & g_1 & g_2 & \cdots & g_{L-1} \end{bmatrix}^T, \tag{4}$$

Then, the power spectra of $g_k$ and $q(t)$ are given by

$$S_g(f) = \left|G(e^{j2\pi f T_0})\right|^2 = \left|\mathbf{v}^H(f, L)\mathbf{g}\right|^2, \tag{5}$$

$$S_q(f) = \left|\int q(t) e^{j2\pi f t} dt\right|^2, \tag{6}$$

where $G(e^{j2\pi f T_0})$ is the frequency response of $\mathbf{g}$.

Substituting (5) and (6) into (2), we have

$$S_p(f) = \left|\mathbf{v}^H(f,L)\mathbf{g}\right|^2 \left|S_q(f)\right|^2, \tag{7}$$

Including the effect of transmitter RF components, the power spectrum of the transmitted waveform can be rewritten as

$$S_{tr}(f) = S_p(f)|R(f)|^2 = |R(f)|^2 \left|\mathbf{v}^H(f,L)\mathbf{g}\right|^2 S_q(f), \tag{8}$$

where $R(f)$ is the frequency response of the transmitter RF components, including amplifiers, samplers and antennas.

For quantitatively evaluating the spectrum-mask utilization efficiency, the normalized effective signal power (NESP) [7] is defined as the ratio of the power of the synthesized UWB waveform to the total power permissible under the spectrum mask. When the spectrum mask is given, the total transmission power allowable under the mask is fixed, so maximization of the transmitted signal power is equivalent to maximization of the NESP.

Let $M(f)$ represent the power spectrum mask, then the NESP maximization problem can be stated as: given $L$, $T_0$, $S_q(f)$ and $M(f)$, find the optimal filter coefficient vector $\mathbf{g}$ that maximizes the total energy in the ultra wideband $\int_{F_p} S_{tr}(f)df$, where $S_{tr}(f)$ is given in (7) and $F_p$ denotes the integration region from 3.1GHz to 10.6GHz [3], subject to the spectral mask constraint that $S_{tr}(f) \leq M(f)$. It can thus be formulated as

$$\max_{\mathbf{g}} \left( \int_{F_p} |R(f)|^2 \left|\mathbf{v}^H(f,L)\mathbf{g}\right|^2 S_q(f)df \right), \tag{9a}$$

$$s.t.\ |R(f)|^2 \left|\mathbf{v}^H(f,L)\mathbf{g}\right|^2 S_q(f) \leq M(f), f \in \left[0, \frac{1}{2T_0}\right], \tag{9b}$$

Compared with the optimization model in [7], the advantage of (9a) and (9b) is that they include the information of transmitter RF component effects $R(f)$.

The cost function in (9a) is a convex quadratic function of $\mathbf{g}$, but since it is to be maximized under a cone constraint, (9a) and (9b) represent a nonconvex optimization problem [7]. In order to prevent getting stuck at a local optimum point, we transform the optimization problem to be

convex as follows:

$$\max_{\mathbf{r}} \left( \mathbf{A}^T \mathbf{r} \right), \tag{10a}$$

$$s.t.\ 0 \leq \mathbf{w}^T(f,L)\mathbf{r} \leq \frac{M(f)}{|R(f)|^2 S_q(f)}, f \in \left[0, \frac{1}{2T_0}\right], \tag{10b}$$

where **r** represents the autocorrelation vector of **g** and is given by

$$\mathbf{r} = \left[ g_0^2 + \cdots + g_{L-1}^2, g_0 g_1 + \cdots + g_{L-2}g_{L-1}, \cdots, g_0 g_{L-1} \right]^T, \tag{11}$$

$$\mathbf{A} = \int_{F_p} |R(f)|^2 S_q(f) \mathbf{w}(f,L) df, \tag{12}$$

$$\mathbf{w}(f,L) = \left[1, 2\cos(2\pi f T_0), \cdots, 2\cos(2\pi f (L-1)T_0)\right]^T, \tag{13}$$

(10a) and (10b) represent a semi-infinite linear programming program [9]. As the different types of basis pulse with different parameters has different PSD, which would results in different NESP in the design. Here we resort to the experiment result in [10] to select the best Gaussian pulse with the proper parameters. By defining $S_{q\&rf}(f) = |R(f)|^2 S_q(f)$, the RF information is incorporated, then the method in [7] can be applied directly. using the software cvx [11] to solve (10), the optimal r can be obtained.

Moreover, since the effect of transmitter RF components is included in the optimization problem, possible mismatches with the regulatory spectrum mask caused by RF distortions on the designed UWB waveform can be effectively avoided. Next, we will further include radio propagation effects into the optimization problem for reducing power losses caused by channel attenuation effects.

*B. Radio propagation effects*

UWB signals typically suffer severe channel attenuation effects. In order to reduce power loss caused by deep fading at certain frequency bands, we can design the transmitted signal spectrum to avoid the deeply faded bands, so that the power efficiency can be further improved. Accordingly, we can rewrite (10a) and (10b) respectively as follows

$$\max_{\mathbf{r}} \left( \mathbf{B}^T \mathbf{r} \right), \tag{14a}$$

$$s.t.\ 0 \leq \mathbf{w}^T(f,L)\mathbf{r} \leq \frac{M(f)}{|R(f)|^2 S_q(f)} M_h(f), f \in \left[0, \frac{1}{2T_0}\right], \tag{14b}$$

where

$$\mathbf{B} = \int_{F_p} |R(f)|^2 S_q(f) M_h(f) \mathbf{w}(f,L) df, \tag{15}$$

$M_h(f)$ is the frequency response of the channel, for which a natural choice is the PSD of the channel impulse response $S_h(f)$.

Both transmitter and receiver can perform the waveform design. When it does in the transmitter, the channel information is estimated at the receiver and fed back to the transmitter by a reporting channel as the cognitive radio approach [12]. When the design is done in receiver, the designed FIR filter taps would be sent back to the transmitter by a reporting channel. The reporting channel would take up some communication resource, but as the environment did not change so much and so frequently, the reported channel information would sustain a considerable long period and the performance gain is considerable. Besides; when the design is in transmitter, the channel information can be regarded as aprior knowledge, which can be obtained with the assistance of the positioning and channel database, as the channel fingerprinting based location approach [13].

Once the optimal $\mathbf{r}$ is obtained, the optimal filter coefficient vector $\mathbf{g}$ can be found via spectral factorization [14][ 15][16].

*C. Spectral Factorization*

To solve the FIR filter taps from the autocorrelation $r$, we use the spectral factorization by the Fejer-Riesz theorem. Let $R(z)$ represent the z-transform of $r$ and $X_{mp}(z)$ denote the unique minimum-phase factor of R(z). An efficient method for minimum phase spectral factorization can be obtained [14][15][16].

$\log X_{mp}(z)$ can be formulated as

$$\log X_{mp}(z) = \alpha(z) + j\varphi(z), \tag{16}$$

where $\alpha(z)=(1/2)\log R(z)$ is known. Since $X_{mp}(z)$ is minimum phase, $\log X_{mp}(z)$ is analytic in the region $\{z| z \geq 1\}$ with the power series expansion:

$$\log X_{mp}(z) = \sum_{n=0}^{\infty} a_n z^{-n}, |z| \geq 1, \tag{17}$$

Consequently, for $z = e^{j\omega}$,

$$\alpha(\omega) = \sum_{n=0}^{\infty} a_n \cos \omega n, \tag{18}$$

$$\varphi(\omega) = -\sum_{n=0}^{\infty} a_n \sin \omega n, \tag{19}$$

Obviously we can see from (18) and (19) that $\alpha(\omega)$ and $\varphi(\omega)$ are Hilbert transform pairs. As $\alpha(\omega)$ could be obtained from $R(z)$, we could first find $\varphi(\omega)$ via the Hilbert transform. Then $X_{mp}(z)$ can be determined from (16). Finally, A Fourier transform yields the coefficients of $X_{mp}(z)$, which gives the desired minimum phase FIR filter coefficients.

## IV. SIMULATION

In addition to designs that comply with the FCC mask $M_{FCC}(f)$, we also seek UWB waveform designs that comply with a tighter mask $M_T(f)$, which is given by

$$M_T(f) = \begin{cases} 0\text{dB} & 3.1\text{GHz} \leq f \leq 10.6\text{GHz} \\ -40\text{dB} & 0 \leq f < 3.1\text{GHz} \\ -15\text{dB} & f > 10.6\text{GHz} \end{cases}. \tag{20}$$

Enforcing the tighter mask allows some margin for "spectral regrowth" due to nonlinearities of the transmitter RF components [7].

In the simulations, we consider both the tighter mask $M_T(f)$ and the normalized FCC mask. The channel model adopted in the simulations is the NLOS CM4 of the IEEE 802.15.3a channel models [17]. The length of FIR filter is set at 30. The basis pulse is the first derivative of the Gaussian pulse with the shape factor being 0.10ns [4]. For simplicity, we let the $R(f) = j2\pi f$, which is commonly known as the derivative effect of a transmitting antenna [4]. The sampling frequency is set to be $f_s = 1/T_0 = 28$GHz.

Fig. 1 shows the transmitted and received spectrum of the waveform designed using (10a) and (10b). Fig. 2 shows the transmitted and received spectrum of the waveform designed using (14a) and (14b). By comparing these two figures, we can see that the waveforms in Fig. 2 incorporate the radio channel information to form notches which would avoid the energy waste in the radio propagation.

For a quantitative evaluation, we define the transmitted power utility ratio at the receiver as

$$\eta = \frac{\int_{F_p} S_{tr}(f) S_h(f) df}{\int_{F_p} S_{tr}(f) df} \times 100\%. \tag{21}$$

We further define $\eta_{ave1}$ and $\eta_{ave2}$ as average values of $\eta$ corresponding to filter coefficients designed with (10) and (14), respectively, averaged over 100 realizations of the channel impulse response. Then, the performance gain $\beta$ can be defined as $\beta = (\eta_{ave2} - \eta_{ave1})/\eta_{ave1} \times 100\%$. Corresponding values of $\eta_{ave1}$, $\eta_{ave2}$ and $\beta$ for the four different channel models (CM1 to CM4) of [17] are presented in Table I. We observe that the transmitted power utility ratio is considerable for every channel model. They have only about 20% of the transmitted power arriving at the receivers. However, the proposed waveform design can let 35% of the transmitted power being received. As the signal to noise ratio (SNR) is a key parameter for the communication performance, when the other situation is the same, the proposed method can save a lot of power. Therefore it is a good candidate for realizing green UWB communication.

## V. CONCLUSION

In this paper, we have proposed a power-efficient UWB waveform design algorithm, targeting at greener wireless communications. Starting from a given regulatory UWB emission mask, the effect of transmitter RF components is included in the UWB waveform design to avoid possible breaking of the spectrum mask at the output of the transmitter, then frequency selective fading is included in the UWB waveform design to enhance the transmitted signal power utility ratio at the receiver. Simulation results have shown that our proposed UWB waveform design algorithm is able to provide good power efficiency while at the same time keeping a high mask utility ratio.

## ACKNOWLEDGMENT

This work was supported in part by the National Natural Science Foundation of China under grant 60772146, the National High Technology Research and Development Program of China (863 Program) under grant 2008AA12Z306 and in part by Science Foundation of Ministry of Education of China under grant 109139.

TABLE I
PERFORMANCE WITH CHANNEL ATTENUATION MASK $S_h(f)$

|  | CM1 | CM2 | CM3 | CM4 |
| --- | --- | --- | --- | --- |
| $\eta_{ave1}$ | 20.33% | 20.62% | 21.14% | 19.80% |
| $\eta_{ave2}$ | 36.46% | 35.57% | 36.82% | 33.29% |
| $\beta$ | 79.34% | 72.46% | 74.16% | 68.12% |

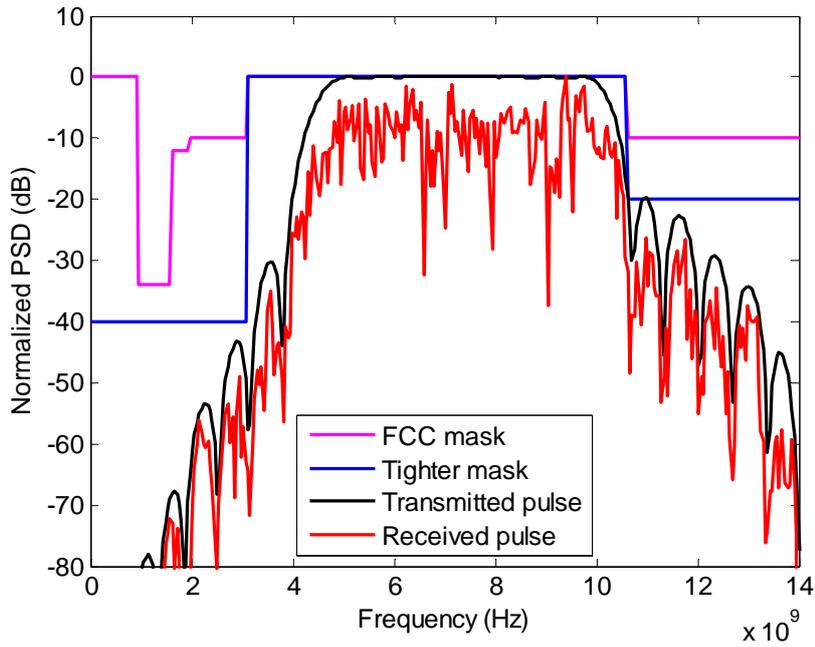

Fig. 1. The transmitted and received spectrum of the waveform designed using (10).

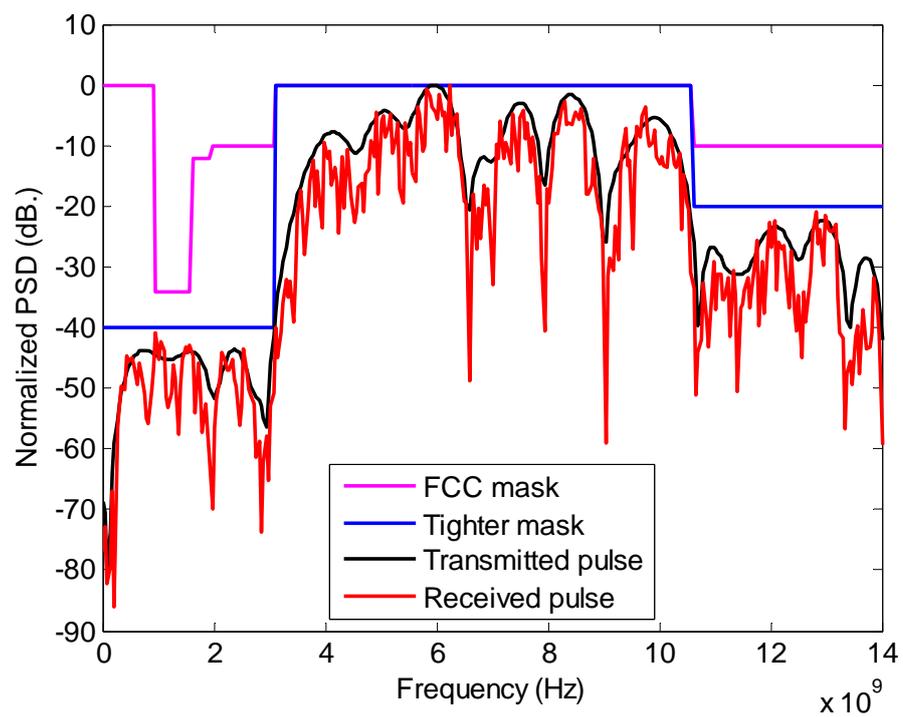

Fig. 2. The transmitted and received spectrum of the waveform designed using (14).